\documentclass[11pt]{article}

\newcommand{\be}{\begin{equation}}
\newcommand{\ee}{\end{equation}}
\newcommand{\bea}{\begin{eqnarray}}
\newcommand{\eea}{\end{eqnarray}}
\newcommand{\lbl}[1]{\label{eq:#1}}
\newcommand{ \rf}[1]{(\ref{eq:#1})}

\def\theequation{\arabic{section}.\arabic{equation}}

\begin{document}

\begin{titlepage}

\begin{flushright}
CPT-2003/P.4510\\
\end{flushright}

\vspace*{0.2cm}
\begin{center}
{\Large {\bf New nonrenormalization theorems \\[0.5cm]for anomalous three point functions
}}\\[2 cm]
{\bf Marc Knecht}$^a$, {\bf Santiago Peris}$^b$, {\bf Michel
Perrottet}$^a$ {\bf and  Eduardo de Rafael}$^a$\\[1cm]

$^a$  {\it Centre  de Physique Th{\'e}orique\\
       CNRS-Luminy, Case 907\\
    F-13288 Marseille Cedex 9, France}\\[0.5cm]
$^b$ {\it Grup de F{\'\i}sica Te{\`o}rica and IFAE\\ Universitat
Aut{\`o}noma
de Barcelona\\  08193 Barcelona, Spain}\\

\end{center}

\vspace*{1.0cm}

\begin{abstract}
Nonrenormalization theorems involving the transverse, i.e. non anomalous, part of the
$\langle VVA \rangle$ correlator in perturbative QCD are proven. Some of the consequences
and questions they raise are discussed.
\end{abstract}

\end{titlepage}

\section{Introduction}
\lbl{int}

\noindent Since their discovery more than thirty years ago \cite{Adler69,BellJackiw69},
anomalous Ward identities for three point functions involving  vector and axial currents
have been studied quite extensively and from various points of view \cite{Zakharov} (see
also Refs. \cite{Treiman85} and \cite{Bertlmann96}). In QCD with three massless flavours,
these anomalous contributions appear in the Green's functions involving the conserved
Noether currents of the global $SU(3)_L\times SU(3)_R$ chiral symmetry.
%
%
Most remarkable in this context is the property that the expression of these anomalous
contributions is protected from perturbative QCD corrections, so that it takes the same
form as in a theory of free quarks
\cite{AdlerBardeen69,Brandeis70,Zee72,LowensteinSchroer73}. Strictly speaking, no similar
statement is available at the nonperturbative level. However, the argument of Ref.
\cite{Witten83} requires that, in an appropriate normalization, the coefficient of the
anomaly be an integer, $N_C$ in the case of QCD. This makes it very likely that the
anomaly is also preserved after non perturbative corrections. Assuming that QCD confines,
and that confinement does indeed not modify the coefficient of the anomaly, then entails
that the $SU(3)_L\times SU(3)_R$ chiral symmetry of QCD is necessarily spontaneously
broken towards its diagonal $SU(3)_V$ subgroup
\cite{tHooft79,Frishman80,ColemanGrossman82}.

These properties are well established and concern the longitudinal part of the
corresponding Green's functions, which is thus completely fixed by the Adler-Bell-Jackiw
anomaly. The transverse parts however are  not affected, and satisfy the naive Ward
identities. Therefore, the \emph{new} result that
 perturbative QCD corrections to the free quark
contribution  are also absent in the \emph{transverse} part of the $\langle VVA\rangle$
three-point function  came as rather unexpected \cite{Vainshtein02,Czarneckietal02}.
Although this was obtained only in a very specific kinematical limit, as we shall see
later on, it is also true even in a more general setting.

The interest in this QCD correlator stems from the fact that it appears in the
determination of the hadronic contribution to a class of two loop weak corrections to the
anomalous magnetic moment of the muon. As has been first emphasized in
Ref.~\cite{Peris:1995bb}, the contribution from the light $u$, $d$ and $s$ quarks to the
transverse part of the $<VVA>$ triangle involves properties of QCD of a nonperturbative
nature. A detailed evaluation, within the framework of large-$N_c$ QCD, has been
published in ref.~\cite{KPPdR02}. However, a recent reanalysis by the authors of
Ref.~\cite{Czarneckietal02} finds also a perturbative contribution to this transverse
part, which originates in a discrepancy between the treatment of the Operator Product
Expansion done in ~\cite{KPPdR02} and the one in \cite{Czarneckietal02}. Although this
discrepancy results in a numerical effect which is too small to influence the present
comparison between theory and experiment in the muon $g-2$, the underlying theoretical
issues involved in this discrepancy are of theoretical interest and, therefore, deserve
special attention.

\indent

\noindent Let us consider the theorem on the non-renormalization of the transverse parts
of the $\langle VVA\rangle$ Green's function discovered in \cite{Vainshtein02}. Actually,
what we shall find is that there is a whole class of theorems, of which the one
considered in Ref.~\cite{Vainshtein02} represents a special case. We shall also provide
an explanation as to the origin and interpretation of these results. Before proceeding,
let us recall a few useful facts concerning anomalous Ward Identities (see also the
discussion in Ref. \cite{GL}).

The Ward identities describing the invariance properties of
massless QCD under the transformations of the global
$SU(3)_L \times SU(3)_R$ chiral symmetry group are
most conveniently
obtained upon considering the transformation properties
of the generating functional $Z[v,a]$, defined as
\bea
e^{iZ[v,a]} &=& \int {\cal D}[G]{\cal D}[\psi,{\overline{\psi}}]
e^{\,i\int d^4 x {\cal L}_{\tiny{\mbox{QCD}}}(v,a)}
\nonumber\\
&=& \int {\cal D}[G] e^{i\int d^4 x {\cal L}_{YM}}\,{\mbox{det}}\,D \,, \eea with
$D=\gamma^\mu(\partial_\mu - i G_\mu - iv_\mu - i\gamma_5 a_\mu)$ the Dirac operator in
presence of the external vector and axial sources for the $SU(3)$ flavour
currents~\footnote{We restrict ourselves to nonsinglet flavour currents, i.e. the sources
$v_\mu$ and $a_\mu$ are singlets in colour space and traceless matrices in flavour
space.}, and in the presence of the gluon field configuration $G_\mu$. The determinant of
the Dirac operator that appears, after integration over the quark fields, in the second
expression for $Z[v,a]$ needs to be regularized. The $\zeta$-function
technique~\footnote{In order to apply this type of regularization, one first has to
perform a rotation to Euclidean space. We assume that this has been done, but we give the
resulting expressions after having rotated back to Minkowski space.} offers a convenient
regularization, \be \ln {\mbox{det}}_{\mu} \,D \,=\,-\,\frac{1}{2}\,\frac{d}{ds}
\,\frac{\mu^{2s}}{\Gamma(s)} \,\int_0^\infty d\lambda \lambda^{s-1}{\mbox{Tr}}
\,e^{-\lambda{\overline D}^2} \bigg\vert_{s=0} \,, \ee with ${\overline D}=\gamma_5 D$.
Other regularizations of the functional determinant, det $D$, may be considered. They
differ from the previous one by a local functional of the sources of dimension less than
or equal to four, \be \ln {\mbox{det}}_F \,D \,=\, \ln {\mbox{det}}_{\mu}\, D \,-\, \int
d^4x F(v,a) \,. \ee Equivalently, these changes of the regularization may be considered
as modifications, by local counterterms, of the chronological products of the $SU(3)_L
\times SU(3)_R$ currents.
%
%
Two particular definitions of the fermion determinant are
of interest.
The first one, $\ln {\mbox{det}}_{AB}\,D$, reproduces the
Adler-Bardeen form of the anomaly \cite{Bardeen69}
and corresponds to the choice
\be
F_{AB}(v,a) \,=\,-i\,\frac{N_C}{12\pi^2}\,{\mbox{tr}}_f\bigg\{
\frac{1}{2}\,\nabla_\mu a_\nu \nabla^\mu a^\nu
\,-\,3iv^{\alpha\beta}a_\alpha a_\beta \,-\,
a_\alpha a_\beta a^\alpha a^\beta \,+\,
2a_\alpha a^\alpha a_\beta a^\beta\bigg\}
\,.
\ee
The second definition reproduces a factorized
or left-right symmetric form of the anomaly \cite{Bardeen69},
$\ln {\mbox{det}}_{\mbox{\scriptsize{LR}}}\,D$,
and corresponds to the choice
\be
F_{\mbox{\scriptsize{LR}}}(v,a) \,=\, F_{AB}(v,a)
\,-\, \frac{N_C}{24\pi^2}\,\epsilon^{\alpha\beta\mu\nu}\,
{\rm{tr}}_f\bigg\{iv_{\mu\nu}\{ v_\alpha,a_\beta\}\,-\,
a_\alpha a_\beta a_\mu v_\nu \,-\,
v_\alpha v_\beta v_\mu a_\nu
\bigg\}
\,.
\ee

\indent

\noindent
Naively, one would expect $Z[v,a]$ to be invariant,
$Z[v,a]= Z[v+\delta v,a+\delta a]$, under the following
variations of the sources
\bea
\delta v_\mu &=& \partial_\mu \alpha \,+\, i[\alpha,v_\mu ] \,+\,
i[\beta,a_\mu]
\nonumber\\
\delta a_\mu &=& \partial_\mu \beta \,+\, i[\alpha,a_\mu ] \,+\,
i[\beta,v_\mu]
\,,
\nonumber\\
\eea
where the Lie algebra valued functions $\alpha(x)=\alpha^a(x)\lambda^a/2$
and $\beta(x)=\beta^a(x)\lambda^a/2$ correspond to infinitesimal
vector and axial transformations, respectively.
Indeed, this specific variation of the sources can in principle
be compensated by a change of variables in the quark fields
corresponding to a local $SU(3)_L \times SU(3)_R$ gauge
transformation. However, the functional measure of integration
${\cal D}[\psi,{\overline\psi}]$ over the quark fields
is not invariant \cite{Fujikawa79,Fujikawa80} under such a
change of variables.
Instead, there appears a non trivial Jacobian, which leads to
\be
\delta \ln {\mbox{det}}_{AB}\,D \,=\,-i\,\frac{N_C}{(4\pi)^2}\,
\int d^4x {\mbox{tr}}_f\, (\beta\Omega)
\,,
\ee
with
\bea
\Omega &=& \epsilon^{\alpha\beta\mu\nu}\bigg[
v_{\alpha\beta}v_{\mu\nu}\,+\,\frac{4}{3}\,
\nabla_\alpha a_\beta \nabla_\mu a_\nu
\,+\,\frac{2i}{3}\,\{v_{\alpha\beta},a_\mu a_\nu\}
\nonumber\\
&&
\,+\,\frac{8i}{3}\,a_\mu v_{\alpha\beta} a_\nu
\,+\,\frac{4}{3}\,a_\alpha a_\beta a_\mu a_\nu
\bigg]
\,,
\eea
and
\be
\delta \ln {\mbox{det}}_{\mbox{\scriptsize{LR}}}\,D \,=\,
-i\,\frac{N_C}{(4\pi)^2}\,\int d^4x {\rm{tr}}_f [
(\alpha + \beta)A(F^R)\,+\,(\alpha - \beta)A(F^L)]
\,,
\ee
where $F^R_\mu = (v+a)_\mu$, $F^L_\mu = (v-a)_\mu$, and
\be
A(F) \,=\,\frac{1}{3}\,\epsilon^{\alpha\beta\mu\nu}\bigg[
2\partial_\alpha F_\beta
\partial_\mu F_\nu \,-\,i\partial_\alpha (F_\beta F_\mu F_\nu)
\bigg]
\,.
\ee

\indent

\noindent

\section{Non renormalization theorems for $\langle VVA\rangle$}
\setcounter{equation}{0}

Let us now consider, for $a,b,c=3,8$, the QCD three point functions \bea
\label{Greencalw}{\cal W}^{abc}_{\mu\nu\rho}(q_1,q_2) &=& i\int d^4x_1 d^4x_2
\,e^{i(q_1\cdot x_1 + q_2\cdot x_2)}
\,\times\,\langle\,0\,\vert\,\mbox{T}\{V^a_\mu(x_1)V^b_\nu(x_2)A^c_\rho(0)\}
\,\vert\,0\,\rangle
\nonumber\\
&\equiv& \frac{1}{2}\,d^{abc}\,{\cal W}_{\mu\nu\rho}(q_1,q_2)\,, \eea of the colour
singlet light flavour currents \be V^a_{\mu} \,=\,
{\overline\psi}\gamma_\mu\,\frac{\lambda^a}{2}\psi \quad ,\quad A^{a}_{\mu} \,=\,
{\overline\psi}\gamma_\mu\gamma_5\,\frac{\lambda^a}{2}\psi \quad ,\quad \psi = \left(
\begin{array}{c} u \\ d \\ s \end{array} \right) \,. \ee
Taking $q_1$ or $q_2$ to be the small momentum going through an external electromagnetic
field, this Green's function appears in the hadronic electroweak contributions to the
muon $g-2$ at the two-loop level \cite{KPPdR02,Czarneckietal02}. To be more precise, in
the muon $g-2$ there is also the contribution from the flavour singlet component of the Z
boson, but we have not included this flavour singlet in the definition  of the Green's
function (\ref{Greencalw}) in an attempt to simplify the discussion which will follow and
because it does not play any crucial role.

Besides ${\cal W}_{\mu\nu\rho}$ in Eq. (\ref{Greencalw}), we shall also need \bea
{\Omega}^{abc}_{\mu\nu\rho}(q_1,q_2) &=& i\int d^4x_1 d^4x_2 \,e^{i(q_1\cdot x_1 +
q_2\cdot x_2)}
\,\times\,\langle\,0\,\vert\,\mbox{T}\{L^a_\mu(x_1)V^b_\nu(x_2)R^c_\rho(0)\}
\,\vert\,0\,\rangle
\nonumber\\
&\equiv& \frac{1}{2}\,d^{abc}\,{\Omega}_{\mu\nu\rho}(q_1,q_2)\,, \lbl{omegadef} \eea with
\be L^a_{\mu} \,=\, {\overline{\psi_L}}\gamma_\mu\,\frac{\lambda^a}{2}\psi_L \quad ,\quad
R^{a}_{\mu} \,=\, {\overline{\psi_R}}\gamma_\mu\,\frac{\lambda^a}{2}\psi_R \quad ,\quad
\psi_{L,R} = \frac{1}{2}(1 \mp \gamma_5) \psi \,. \ee

The two Green's functions ${\cal W}_{\mu\nu\rho}$ and ${\Omega}_{\mu\nu\rho}$ are
actually related. Use of charge-conjugation invariance allows one to do away with the
combinations $<VVV>$ and $<VAA>$ in the function $\Omega_{\mu\nu\rho}$ and obtain that,
in fact, \be \label{omegaw}{\Omega}_{\mu\nu\rho}(q_1,q_2)\,=\,\frac{1}{4}\, \bigg[ {\cal
W}_{\mu\nu\rho}(q_1,q_2)\,-\,{\cal W}_{\rho\nu\mu}(-q_1-q_2,q_2) \bigg] \,. \ee In the
above definitions, the time ordering corresponds to the definition of the chronological
product T which preserves the conservation of the vector currents, i.e. to the
prescription ${\rm{det}}_{AB}\,D$. If instead we wish to use the prescription
${\rm{det}}_{\small{\rm{LR}}}\,D$, the corresponding chronological product will be
denoted by ${\widehat{\mbox{T}}}$. For instance, \bea\lbl{formalomegahat}
{\widehat\Omega}^{abc}_{\mu\nu\rho}(q_1,q_2) &=& i\int d^4x_1 d^4x_2 \,e^{i(q_1\cdot x_1
+ q_2\cdot x_2)}
\,\times\,\langle\,0\,\vert\,{\widehat{\mbox{T}}}\{L^a_\mu(x_1)V^b_\nu(x_2)R^c_\rho(0)\}
\,\vert\,0\,\rangle
\nonumber\\
&\equiv& \frac{1}{2}\,d^{abc}\,{\widehat\Omega}_{\mu\nu\rho}(q_1,q_2)\,,
\eea
and
\bea
{\widehat{\cal W}}^{abc}_{\mu\nu\rho}(q_1,q_2) &=& i\int d^4x_1 d^4x_2
\,e^{i(q_1\cdot x_1 + q_2\cdot x_2)}
\,\times\,\langle\,0\,\vert\,{\widehat{\mbox{T}}}\{V^a_\mu(x_1)V^b_\nu(x_2)A^c_\rho(0)\}
\,\vert\,0\,\rangle
\nonumber\\
&\equiv& \frac{1}{2}\,d^{abc}\,{\widehat{\cal W}}_{\mu\nu\rho}(q_1,q_2)\,. \eea The
relation between ${\widehat\Omega}_{\mu\nu\rho}$ and ${\widehat{\cal W}}_{\mu\nu\rho}$ is
the same as the one between ${\Omega}_{\mu\nu\rho}$ and ${{\cal W}}_{\mu\nu\rho}$.

Furthermore,

\bea {\cal W}_{\mu\nu\rho}^{abc}(q_1,q_2) &=& {\widehat{\cal
W}}_{\mu\nu\rho}^{abc}(q_1,q_2)
\nonumber\\
&&\quad +\,i \int d^4x_1 d^4x_2 d^4z \,e^{i(q_1\cdot x_1 + q_2\cdot x_2)}
\,\frac{1}{i^3}\,
\,\frac{\delta^3 [F_{\mbox{\scriptsize{LR}}}(v,a) - F_{AB}(v,a)](z)} {\delta
v^{a\mu}(x_1) \delta v^{b\nu}(x_2) \delta a^{c\rho}(0)} \bigg\vert_{v=a=0} \,, \nonumber
\eea which gives \be \mathcal{W}_{\mu\nu\rho}(q_1,q_2) \,=\, {\widehat{\cal
W}}_{\mu\nu\rho}(q_1,q_2) \,+\,\frac{N_C}{12\pi^2}\,\epsilon_{\mu\nu\rho\alpha}(q_1 -
q_2)^\alpha \,, \ee

\noindent whereas
\begin{equation}\label{shift}
    \Omega_{\mu\nu\rho}(q_1,q_2)=\widehat{\Omega}_{\mu\nu\rho} (q_1,q_2)
    - \frac{N_C}{16 \pi^2}\ \epsilon_{\mu\nu\rho\tau}\ q_{2}^{\tau}\ .
\end{equation}

\noindent The Ward identities satisfied by these correlators  read (we work in the chiral
limit), \bea \{q_1^\mu\ ;\ q_2^\nu\}\,{{\cal W}}_{\mu\nu\rho}(q_1,q_2) &=&\{0\ ;\ 0\}
\nonumber\\
\ (q_1 +q_2)^\rho\,{{\cal W}}_{\mu\nu\rho}(q_1,q_2) &=&
-\, \frac{N_C}{4\pi^2}\,\epsilon_{\mu\nu\sigma\tau}q_1^\sigma q_2^\tau
\,,
\eea
\bea
q_1^\mu\,{{\Omega}}_{\mu\nu\rho}(q_1,q_2) &=&
-\,\frac{N_C}{16\pi^2}\,\epsilon_{\mu\nu\rho\tau}q_1^\mu q_2^\tau
\nonumber\\
q_2^\nu\,{{\Omega}}_{\mu\nu\rho}(q_1,q_2) &=& 0
\nonumber\\
\ (q_1 +q_2)^\rho\,{{\Omega}}_{\mu\nu\rho}(q_1,q_2) &=& -\,
\frac{N_C}{16\pi^2}\,\epsilon_{\mu\nu\sigma\tau}q_1^\sigma q_2^\tau \,, \eea whereas \be
\{q_1^\mu\ ;\ q_2^\nu\ ;\ (q_1+q_2)^\rho\}\,{{\widehat\Omega}}_{\mu\nu\rho}(q_1,q_2)
\,=\,\{0\ ;\ 0\ ;\ 0\} \,. \ee

\indent

\noindent Let us now discuss the transformation properties of these various quantities
under the action of the $SU(3)_L \times SU(3)_R$ flavour group. The ordinary product
$L^a(x)V^b(y)R^c(0)$ transforms under the representation ({\bf 8}$_L\otimes${\bf 8}$_L$\
,\ {\bf 8}$_R$)$\ \oplus \ $({\bf 8}$_L$\ ,\ {\bf 8}$_R\otimes${\bf 8}$_R$), which does
not project onto the singlet representation ({\bf 1}$_L$,{\bf 1}$_R$). The same property
is still true for the canonical chronological product of these currents, defined with the
help of the step function in the time variables. However, this does not lead to a
covariant time ordering. On the other hand, a covariant chronological product is likely
to introduce $SU(3)_L \times SU(3)_R$ invariant contributions. This is precisely not the
case for the chronological product ${\widehat{\mbox{T}}}$, which rests on the
prescription ${\rm{det}}_{\mbox{\scriptsize{LR}}}$ for the fermionic determinant. Indeed,
with this choice of time ordering, the three point function
${\widehat\Omega}_{\mu\nu\rho}(q_1,q_2)$ satisfies the naive Ward identities. Anomalous
contributions then occur only in the Ward identities for the three or four point
functions involving only left currents or only right currents. Therefore,
${\widehat{\mbox{T}}}\{L^a(x)V^b(y)R^c(0)\}$ has the same transformation properties as
the ordinary product, and thus ${\widehat\Omega}_{\mu\nu\rho}(q_1,q_2)$ is an order
parameter of the $SU(3)_L \times SU(3)_R$ chiral symmetry, which means that it does not
receive perturbative QCD corrections at any order, i.e. \be
{\widehat\Omega}_{\mu\nu\rho}(q_1,q_2)\vert_{\mathrm{pQCD}}\,=\,0 \,. \lbl{pQCD} \ee

\indent

\noindent The Ward identities restrict the general decomposition of ${{\cal
W}}_{\mu\nu\rho}(q_1,q_2)$ into invariant functions to four terms

\bea \label{calw}{{\cal W}}_{\mu\nu\rho}(q_1,q_2) &=&
 -\,\frac{1}{8\pi^2}\,\bigg\{
-w_L\left(q_1^2,q_2^2,(q_1+q_2)^2\right)\,(q_1+q_2)_\rho\, \epsilon_{\mu\nu\alpha\beta}\
q_1^\alpha q_2^\beta
\quad\quad\quad\quad\quad\quad\quad\nonumber\\
&&\quad\quad\quad\quad +\,
w_T^{(+)}\left(q_1^2,q_2^2,(q_1+q_2)^2\right)\,t^{(+)}_{\mu\nu\rho}(q_1,q_2)
\nonumber \\
 &&\quad\quad\quad\quad\quad\quad +\,w_T^{(-)}
 \left(q_1^2,q_2^2,(q_1+q_2)^2\right)\,t^{(-)}_{\mu\nu\rho}(q_1,q_2)
 \nonumber\\ && \quad\quad\quad\quad\quad\quad\quad\quad+\,
{\widetilde{w}}_T^{(-)}\left(q_1^2,q_2^2,(q_1+q_2)^2\right)\,{\widetilde{t}}^{(-)}_{\mu\nu\rho}(q_1,q_2)
\bigg\}\,, \eea with the transverse tensors \bea t^{(+)}_{\mu\nu\rho}(q_1,q_2) &=&
q_{1\nu}\,\epsilon_{\mu\rho\alpha\beta}\ q_1^\alpha q_2^\beta \,-\,
q_{2\mu}\,\epsilon_{\nu\rho\alpha\beta}\ q_1^\alpha q_2^\beta \,-\, (q_{1}\cdot
q_2)\,\epsilon_{\mu\nu\rho\alpha}\ (q_1 - q_2)^\alpha
\nonumber\\
&& \quad\quad+\ \frac{q_1^2 + q_2^2 - (q_1+q_2)^2}{(q_1 + q_2)^2}\
\epsilon_{\mu\nu\alpha\beta}\ q_1^\alpha q_2^\beta(q_1 + q_2)_\rho
\nonumber \ , \\
t^{(-)}_{\mu\nu\rho}(q_1,q_2) &=& \left[ (q_1 - q_2)_\rho \,-\, \frac{q_1^2 - q_2^2}{(q_1
+ q_2)^2}\,(q_1 + q_2)_\rho \right] \,\epsilon_{\mu\nu\alpha\beta}\ q_1^\alpha q_2^\beta
\nonumber\\
{\widetilde{t}}^{(-)}_{\mu\nu\rho}(q_1,q_2) &=& q_{1\nu}\,\epsilon_{\mu\rho\alpha\beta}\
q_1^\alpha q_2^\beta \,+\, q_{2\mu}\,\epsilon_{\nu\rho\alpha\beta}\ q_1^\alpha q_2^\beta
\,-\, (q_{1}\cdot q_2)\,\epsilon_{\mu\nu\rho\alpha}\ (q_1 + q_2)^\alpha \,. \lbl{tensors}
\eea Bose symmetry entails \bea w_T^{(+)}\left(q_2^2,q_1^2,(q_1+q_2)^2\right) &=& +
w_T^{(+)}\left(q_1^2,q_2^2,(q_1+q_2)^2\right)
\nonumber\\
w_T^{(-)}\left(q_2^2,q_1^2,(q_1+q_2)^2\right) &=& -
w_T^{(-)}\left(q_1^2,q_2^2,(q_1+q_2)^2\right)
\nonumber\\
{\widetilde{w}}_T^{(-)}\left(q_2^2,q_1^2,(q_1+q_2)^2\right) &=& -
{\widetilde{w}}_T^{(-)}\left(q_1^2,q_2^2,(q_1+q_2)^2\right) \,. \eea In addition, the
longitudinal part is entirely fixed by the anomaly, \be
w_L\left(q_1^2,q_2^2,(q_1+q_2)^2\right)\,=\, - \frac{2N_C}{(q_1 + q_2)^2}\,.
\lbl{w_Lexpr} \ee A straightforward computation then leads to \bea
{\widehat\Omega}_{\mu\nu\rho}(q_1,q_2) &=& -\,\frac{1}{32\pi^2}\,\Bigg\{
-\epsilon_{\mu\nu\rho\alpha}\ q_2^{\alpha}\, \left[ q_1^2\ w_L\left((q_1 +
q_2)^2,q_2^2,q_1^2\right) \,+\, 2N_C \right] +
\nonumber\\
&&
\!\!\!\!\!\!\!\!\!\!\!\!\!\!\!\!\!\!\!\!\!\!\!\!\!\!\!\!\!\!\!\!\!\!\!\!\!\!\!\!\!\!\!\!+\,\left[
q_1^2\ w_L\left((q_1 + q_2)^2,q_2^2,q_1^2\right) \,-\, (q_1 + q_2)^2\
w_L\left(q_1^2,q_2^2,(q_1+q_2)^2\right) \right]\ \frac{(q_1+q_2)_\rho}{(q_1 + q_2)^2}\,
\,\,\epsilon_{\mu\nu\alpha\beta}\ q_1^\alpha q_2^\beta
\nonumber\\
&&
\!\!\!\!\!\!\!\!\!\!\!\!\!\!\!\!\!\!\!\!\!\!\!\!\!\!\!\!\!\!\!\!\!\!\!\!\!\!\!\!\!\!\!\!+
t^{(+)}_{\mu\nu\rho}(q_1,q_2) \,\bigg[ w_T^{(+)}\left(q_1^2,q_2^2,(q_1+q_2)^2\right)
-\,\frac{1}{2}\,w_L\left((q_1 + q_2)^2,q_2^2,q_1^2\right)
\nonumber\\
&& +\,\frac{(q_2^2 \,+\, q_1\cdot q_2)}{q_1^2}\ w_T^{(+)}\left((q_1 +
q_2)^2,q_2^2,q_1^2\right) -\, \left( \frac{q_1\cdot q_2}{q_1^2}\,+\,
1\right)\,w_T^{(-)}\left((q_1 + q_2)^2,q_2^2,q_1^2\right) \bigg]
\nonumber\\
&&
\!\!\!\!\!\!\!\!\!\!\!\!\!\!\!\!\!\!\!\!\!\!\!\!\!\!\!\!\!\!\!\!\!\!\!\!\!\!\!\!\!\!\!+
t^{(-)}_{\mu\nu\rho}(q_1,q_2) \,\bigg[ w_T^{(-)}\left(q_1^2,q_2^2,(q_1+q_2)^2\right)
+\,\frac{1}{2}\,w_L\left((q_1 + q_2)^2,q_2^2,q_1^2\right)
\nonumber\\
&&  -\,\frac{(q_1^2 \,+\, q_2^2 \,+\, q_1\cdot q_2)}{q_1^2}\ w_T^{(+)}\left((q_1 +
q_2)^2,q_2^2,q_1^2\right) +\, \frac{q_1\cdot q_2}{q_1^2}\,w_T^{(-)}\left((q_1 +
q_2)^2,q_2^2,q_1^2\right) \bigg]
\nonumber\\
&&\!\!\!\!\!\!\!\!\!\!\!\!\!\!\!\!\!\!\!\!\!\!\!\!\!\!\!\!\!\!\!\!\!\!\!\!\!\!\!\!\!\!\!
+ {\widetilde{t}}^{(-)}_{\mu\nu\rho}(q_1,q_2) \,\bigg[
{\widetilde{w}}_T^{(-)}\left(q_1^2,q_2^2,(q_1+q_2)^2\right)
\,+\,{\widetilde{w}}_T^{(-)}\left((q_1 + q_2)^2,q_2^2,q_1^2\right)
-\,\frac{1}{2}\,w_L\left((q_1 + q_2)^2,q_2^2,q_1^2\right)
\nonumber\\
&& \!\!\!\!\!\!\!\!\!\!\!\!\!\!\! +\,\frac{(q_1^2 \,+\, q_2^2 \,+\, q_1\cdot
q_2)}{q_1^2}\, w_T^{(+)}\left((q_1 + q_2)^2,q_2^2,q_1^2\right) -\, \left( \frac{q_1\cdot
q_2}{q_1^2}\,-\,
1\right)\,w_T^{(-)}\left((q_1 + q_2)^2,q_2^2,q_1^2\right) \bigg]\Bigg\}\ .\nonumber \\
\lbl{Omegahat} \eea

\indent

\noindent Due to the expression \rf{w_Lexpr} of
$w_L\left(q_1^2,q_2^2,(q_1+q_2)^2\right)$, the two first terms on the right-hand side of
Eq. \rf{Omegahat} vanish identically. Since the three tensors \rf{tensors} are
independent, the property \rf{pQCD} implies that the combinations of invariant functions
that multiply $t^{(+)}_{\mu\nu\rho}(q_1,q_2)$, $t^{(-)}_{\mu\nu\rho}(q_1,q_2)$ and
${\widetilde{t}}^{(-)}_{\mu\nu\rho}(q_1,q_2)$ in the expression \rf{Omegahat} have to
vanish to all orders in perturbation theory. Consequently, the following three non
renormalization theorems follow, \bea &&\!\!\!\!\!\!\!\!\!\! \Biggl\{\left[w_T^{(+)}
\,+\, w_T^{(-)}\right] \left(q_1^2,q_2^2,(q_1+q_2)^2\right) \, -\, \left[w_T^{(+)} \,+\,
w_T^{(-)}\right] \left((q_1 + q_2)^2,q_2^2,q_1^2\right)\Biggr\}_{\mathrm{pQCD}}\ =\ 0
\nonumber\\
\\
&&\!\!\!\!\!\!\!\!\!\! \Biggl\{\left[{\widetilde{w}}_T^{(-)} \,+\, w_T^{(-)}\right]
\left(q_1^2,q_2^2,(q_1+q_2)^2\right) \, +\, \left[{\widetilde{w}}_T^{(-)} \,+\,
w_T^{(-)}\right] \left((q_1 + q_2)^2,q_2^2,q_1^2\right)\Biggr\}_{\mathrm{pQCD}}\  =\  0
\nonumber\\
\eea and \bea \label{theorems}&& \!\!\!\!\!\!\!\!\!\!\!\!\!\!\Biggl\{\left[w_T^{(+)}
\,+\, {\widetilde{w}}_T^{(-)}\right]\left(q_1^2,q_2^2,(q_1+q_2)^2\right) \,+\, \left[{{w}}_T^{(+)}
\,+\, {\widetilde{w}}_T^{(-)}\right] \left((q_1 +
q_2)^2,q_2^2,q_1^2\right)\Biggr\}_{\mathrm{\!\!pQCD}} \!\!\!\!\!\!- w_L\left((q_1 +
q_2)^2,q_2^2,q_1^2\right)\nonumber \\ &&  = -\Biggl\{ \frac{2\ (q_2^2 + q_1\cdot
q_2)}{q_1^2}\, w_T^{(+)} \left((q_1 + q_2)^2,q_2^2,q_1^2\right) \ -\, 2\,\frac{q_1\cdot
q_2}{q_1^2}\, w_T^{(-)} \left((q_1 + q_2)^2,q_2^2,q_1^2\right)\Biggr\}_{\mathrm{pQCD}}
\,, \eea involving the transverse part of the $\langle VVA\rangle$ correlator ${{\cal
W}}_{\mu\nu\rho}(q_1,q_2)$, and which hold for all values of the momentum transfers
$q_1^2$, $q_2^2$ and $(q_1 + q_2)^2$. Notice that in Eq. (\ref{theorems}) the
longitudinal function $w_L$ does not need to carry the subindex ``pQCD'' due to the
nonrenormalization of the anomaly.

The non renormalization theorem obtained in Refs. \cite{Vainshtein02,Czarneckietal02}
appears as a particular case. Indeed, upon taking $q_1 = k \pm q$, $q_2 = -k$, and
keeping only the terms linear in the momentum $k$, one readily obtains
\bea\label{tensors}
 t^{(+)}_{\mu\nu\rho}(k\pm q,-k) &=&
q^2 \epsilon_{\mu\nu\rho\sigma}k^\sigma -
q_\mu \epsilon_{\nu\rho\alpha\beta}q^\alpha k^\beta -
q_\rho \epsilon_{\mu\nu\alpha\beta} q^\alpha k^\beta
\,+\,{\cal O}(k^2)
\nonumber\\
t^{(-)}_{\mu\nu\rho}(k\pm q,-k) &=& {\cal O}(k^2)
\nonumber\\
{\widetilde{t}}^{(-)}_{\mu\nu\rho}(k\pm q,-k) &=& q^2 \epsilon_{\mu\nu\rho\sigma}k^\sigma
- q_\mu \epsilon_{\nu\rho\alpha\beta}q^\alpha k^\beta - q_\rho
\epsilon_{\mu\nu\alpha\beta} q^\alpha k^\beta \,+\,{\cal O}(k^2) \,. \eea Within this
same kinematical configuration, the three non renormalization theorems then reduce to one
single equality, namely the result of Refs. \cite{Vainshtein02,Czarneckietal02} \be
w_L(Q^2)\,=\, 2\ w_T(Q^2)_{\mathrm{pQCD}} \,, \lbl{vainshtein} \ee where $Q^2\equiv -q^2$
and \bea \label{connection}w_L(Q^2) &=& w_L(-Q^2,0,-Q^2)
\nonumber\\
w_T(Q^2) &=& {w}_T^{(+)}(-Q^2,0,-Q^2) \,+\,
{\widetilde{w}}_T^{(-)}(-Q^2,0,-Q^2)
\,.
\eea

\indent

\noindent Our result \rf{pQCD} thus contains and extends the non renormalization theorem
of Refs. \cite{Vainshtein02,Czarneckietal02} to general values of the momentum transfers.
More interestingly perhaps, it identifies the origin and the meaning of these results:
they merely follow from the fact that, with an appropriate definition of the
chronological product, the $\langle LLR\rangle$ three point correlator \rf{omegadef} is
an order parameter of the chiral symmetry group of QCD with three massless flavours. The
proof of the particular case \rf{vainshtein} outlined in Ref. \cite{Vainshtein02} relies
on arguments of a diagrammatic kind. In Appendix A, we show that a proof of \rf{pQCD} can
also be established along these lines. The proof is based on the same argumentation as
that of Refs. \cite{AdlerBardeen69,LowensteinSchroer73} in the case of the chiral anomaly
(see also Appendix B).

\section{ The function $w_{L}(Q^2)-2w_{T}(Q^2)$ in full QCD.}
\setcounter{equation}{0}

\noindent Using the general decomposition of $\widehat{\Omega}_{\mu\nu\rho}(q_1,q_2)$ in
eq.~\rf{Omegahat}, we can relate the amplitude $w_{L}(Q^2)-2w_{T}(Q^2)$ to the three
point function $\widehat{\Omega}_{\mu\nu\rho}(q_1,q_2)$ in the appropriate kinematic
configuration where $q_1=k\pm q$, $q_2=-k$, and only the terms linear in $k$ are kept,
namely \be \label{omegahat}\widehat{\Omega}_{\mu\nu\rho}(k\pm
q,-k)=\!\frac{1}{32\pi^2}\left[ w_{L}(Q^2)\!-\!2w_{T}(Q^2)\right] \left(q^2
\!\epsilon_{\mu\nu\rho\sigma}k^{\sigma}\!-\!q_{\mu}\epsilon_{\nu\rho\alpha\beta}
q^{\alpha}k^{\beta}\!-\!q_{\rho}\epsilon_{\mu\nu\alpha\beta}
q^{\alpha}k^{\beta}\right)\!+\!{\cal O}(k^2)\,. \ee The result of Eq.
(\ref{eq:vainshtein}) is nothing but the statement that $\widehat{\Omega}_{\mu\nu\rho} $
vanishes in perturbation theory\footnote{The reader who has found our derivation of Eq.
(\ref{omegahat}) a bit too formal is referred to Appendix B for a less formal version of
it. }. Contracting this expression with $\epsilon^{\mu\nu\rho\lambda}$, we have \be
\lim_{k\to 0}\frac{\partial}{\partial k^{\lambda}}
\left(\epsilon^{\mu\nu\rho\lambda}\widehat{\Omega}_{\mu\nu\rho}(k\pm q,-k)
\right)=\frac{3}{8\pi^2}\ Q^2\left[ w_{L}(Q^2)-2w_{T}(Q^2)\right]\,, \ee or equivalently,
using the definition of $\widehat{\Omega}_{\mu\nu\rho}(q_1,q_2)$ in
eq.~\rf{formalomegahat}, \be\lbl{moment}
 Q^2\left[
w_{L}(Q^2)-2w_{T}(Q^2)\right]=\frac{16\pi^2}{\sqrt{3}} \int\! d^4x\!\!\int\! d^4y\
e^{iq\cdot x} (y\!-\!x)_{\lambda}\epsilon^{\mu\nu\rho\lambda}\langle
0\vert{\widehat{\mbox{T}}} \{L^3_\mu(x)V^3_\nu(y)R^8_\rho(0)\} \vert 0\rangle\,, \ee
explicitly showing the fact that the function $ Q^2\left[ w_{L}(Q^2)-2w_{T}(Q^2)\right]$
is an order parameter of S$\chi$SB at all values of its argument. Obviously, in the
chiral limit, perturbation theory produces a vanishing result for the right-hand side of
Eq. (\ref{eq:moment}) to all orders. Furthermore, since the function $w_L(Q^2)$ is exact
\cite{Bardeen69}, it follows that $w_T(Q^2)$ cannot receive contributions in pQCD beyond
its lowest order value. However, this all-orders result for $w_T(Q^2)$ in perturbation
theory cannot go through at the nonperturbative level. Indeed, the low-$Q^2$ behaviour of
$w_{T}(Q^2)$ is governed by the ${\cal O}(p^6)$ effective chiral Lagrangian in the
odd-parity sector. The relevant coupling is the term (see Ref.~\cite{BGT02} for
notations) \be \label{last}{\cal L}_6^W = C_{22}^W \,\epsilon_{\mu\nu\alpha\beta}\,
\mathrm{Tr}\Big(u^\mu\{\nabla_\gamma f_+^{\gamma\nu},f_+^{\alpha\beta}\}\Big) + \cdots\,,
\ee which fixes $w_{T}(0)$ in terms of the low-energy constant $C_{22}^W$ as follows \be
\label{last2} w_{T}(0) = {128\pi^2}C_{22}^W\,. \ee Unfortunately, there is no model
independent information on this constant. Therefore, contrary to the case of the
anomalous amplitude $w_{L}(Q^2)$ which is fixed by the ${\cal O}(p^4)$ Wess-Zumino term
in the effective chiral Lagrangian, the transverse amplitude $w_{T}(Q^2)$ \emph{has no
pole} at $Q^2= 0$. This clearly shows that $w_{T}(Q^2)$ is affected by {\it
nonperturbative} QCD corrections, in contrast to the case of $w_{L}(Q^2)$ \footnote{The
argument which requires $w_{L}(Q^2)$ to be normalized by an integer \cite{Witten83} also
does not apply to $w_{T}(Q^2)$.}. At large values of $Q^2$ one can start seeing this
different behaviour through the use of the Operator Product Expansion. In this way, both
the authors of Ref. \cite{KPPdR02} and \cite{Czarneckietal02,Vainshtein02} agree that
$w_T(Q^2)$ receives, as the leading nonperturbative contribution at large $Q^2$, a term
proportional to
\begin{equation}\label{nonp}
    \left\{w_T(Q^2)\right\}_{NP}
    \sim \frac{\alpha_s \langle\overline{\psi} \psi\rangle^2}{Q^6 F_{\pi}^2}\ .
\end{equation}
Since the function $w_L(Q^2)$ is exactly given by its perturbative result, i.e.
$2N_C/Q^2$, Eqs.  (\ref{last2}) and (\ref{nonp})  break the perturbative degeneracy
$w_L(Q^2)=2\ w_T(Q^2)$. Notice as well that $w_T(Q^2)$ can only receive nonperturbative
contributions in the Operator Product Expansion coming from operators which are order
parameters of spontaneous chiral symmetry breaking. For instance, a contribution from the
gluon condensate $\frac{<G^2>}{Q^6}$ is excluded. This operator was wrongly allowed in
the analysis of Ref. \cite{Czarneckietal02}, although it was then numerically neglected
on the basis of being accompanied by a one-loop suppressed Wilson coefficient.

 The main
difference between the analysis of Ref. \cite{KPPdR02} and
\cite{Czarneckietal02}\cite{Vainshtein02} lies in the existence (or not) of a $1/Q^2$
contribution to the function $w_T(Q^2)$ at large values of $Q^2$. The analysis of Ref.
\cite{Czarneckietal02, Vainshtein02} was based on a separation of virtuality in momentum
and found a contribution which goes like $w_T(Q^2)=N_C/Q^2$ from the region of high
virtuality in the \emph{perturbative} quark loop. In Ref. \cite{KPPdR02}, on the other
hand, no contribution of $\mathcal{O}(1/Q^2)$ was found. The calculation in
\cite{KPPdR02} was based on the fact that $w_T(Q^2)$ satisfies an unsubtracted dispersion
relation and that the high-energy contribution to  $\mathrm{Im}\ w_T$ from the
perturbative QCD continuum vanishes in the chiral limit \cite{appendix}. Closely related
to this point is the interpretation of the anomaly as a subtraction in the corresponding
dispersion relation for $w_L(Q^2)$\cite{Zakharov}.

As we have discussed after Eq. (\ref{eq:moment}), there is no doubt that
$w_T(Q^2)=N_C/Q^2$ in perturbation theory. The question is whether this is also true  in
the exact theory, for large enough values of $Q^2$. The result of Ref. \cite{KPPdR02}
rests on the assumption that the leading large-$Q^2$ contribution to the function
$w_T(Q^2)$  comes from the region of large momentum in its perturbative imaginary part
(i.e. from the $q\overline{q}$ continuum). This assumption is based on the notion of
(global) duality between quarks and hadrons \cite{Poggio}; notion which is heavily based
on examples such as the process $e^{+}e^{-}\rightarrow\mathrm{hadrons}$.

On the other hand, the analysis of Ref. \cite{Czarneckietal02, Vainshtein02} is based on
a separation of virtuality in momentum in the perturbative quark loop, assuming that this
carries through also at the nonperturbative level. Even though a proof of the Operator
Product Expansion is still lacking in QCD, the $1/Q^2$ contribution obtained in
\cite{Czarneckietal02, Vainshtein02} is difficult to avoid since it originates from the
region of infinitely short distances. This is completely unlike the case of
$e^{+}e^{-}\rightarrow\mathrm{hadrons}$ because this $1/Q^2$ behavior, although being
perturbative, cannot be associated with any $q \overline{q}$ continuum in the chiral
limit and this explains why the $1/Q^2$ was not found in \cite{KPPdR02}. In the light of
this discussion, we are ready to accept the result of \cite{Czarneckietal02,
Vainshtein02} at this point, even though we feel that a better understanding of the role
played by the $q \overline{q}$ continuum would be desirable.

We find, however, that the properties of the function $w_T(Q^2)$ are then rather
intriguing and in the rest of this note we would like to point out several of the
``curiosities'' which, we think, may deserve further investigation. Firstly, upon taking
the limit $q\to 0$ in \rf{moment}, one obtains \be N_C \, =\, \frac{8\pi^2}{\sqrt{3}}
\int\! d^4x\!\!\int\! d^4y\ (y\!-\!x)_{\lambda}\epsilon^{\mu\nu\rho\lambda}\langle
0\vert{\widehat{\mbox{T}}} \{L^3_\mu(x)V^3_\nu(y)R^8_\rho(0)\} \vert 0\rangle\ , \ee
since $w_L(Q^2)$ has a pole at $Q^2=0$ (see Eq. (\ref{eq:w_Lexpr})) but $w_T(Q^2)$ is
regular. We find this result quite amazing as it equates the residue of the anomaly --a
term which can be computed in perturbation theory-- to a Green's function which measures
the spontaneous breakdown of chiral symmetry. It would be very interesting to check this
result by nonperturbative methods such as, for instance, lattice gauge theories.

Related to the previous discussion is the fact that all the hadronic QCD sum rules we
know which originate in pQCD short-distance properties are of the type \be
\int_{0}^{\infty} dt\ \omega(t,M^2) \rho (t)\sim
N_C\left[1+\mathcal{O}\left(\frac{\alpha_s(M^2)}{\pi}\right)\right]\,, \ee where
$\rho(t)$ denotes some generic spectral function and $\omega(t,M^2)$ an appropriate
weight function, e.g., $\omega(t,M^2)=\exp(-t/M^2)$ in the so called Laplace or Borel QCD
sum rules. In all these sum rules, it is the presence of $\alpha_s(M^2)$ corrections
which controls the regime in the euclidean at which we can trust the pQCD calculations.
By contrast, the behaviour \be\lbl{vainpQCD} \lim_{Q^2\rightarrow
\infty}w_{T}(Q^2)=\frac{N_C}{Q^2}\,, \quad {\mbox{\rm with no $\alpha_s(Q^2)$
corrections}}\,, \ee leaves us with no scale to gauge the validity of the asymptotic pQCD
behaviour.

In fact, Eq.~(\ref{eq:vainpQCD}) implies an exact QCD sum rule of a new type:
\be\lbl{sumrule} \int_{0}^{\infty}dt\  \frac{1}{\pi}\mathrm{Im}\  w_{T}(t)= N_C\,, \ee
which follows from the fact that $w_{T}(q^2)$ obeys an unsubtracted dispersion relation.
The sum rule (\ref{eq:sumrule}) thus clearly shows a marked difference with respect to,
say, the first Weinberg sum rule,
\begin{equation}\label{wsr}
    \int_{0}^{\infty}\ dt \ \frac{1}{\pi}
    \big[ \mathrm{Im}\Pi_V(t) - \mathrm{Im}\Pi_A(t) \big] =F^2_{\pi}\,,
\end{equation}
where {\it both sides} receive subleading corrections in the $1/N_C$ expansion. The
striking feature about the sum rule in Eq.~(\ref{eq:sumrule}) is that the r.h.s. is
exact. There are no $1/N_C$ corrections on the r.h.s., while the l.h.s., which is an
integral of hadronic contributions, has certainly subleading terms in the $1/N_C$
expansion; e.g., those generated by multiparticle states. For these subleading terms,
this implies a very curious fine tuning of the various terms on the l.h.s. which have to
add up to zero.

There is also and exact QCD sum rule for the $w_{L}(Q^2)$ amplitude \be\lbl{wlsumrule}
\int_{0}^{\infty}dt\  \frac{1}{\pi}\mathrm{Im}\  w_{L}(t)=2\  N_C\,, \ee which, in the
hadronic spectrum, is fulfilled by the Goldstone pole alone: $\frac{1}{\pi}\mathrm{Im}\
w_{L}(t)=2\ N_C\ \delta(t)\,.$ Therefore, there is nothing surprising about this sum
rule, which just reflects the fact that the anomaly is an exact result. By contrast, the
sum rule in Eq.~\rf{sumrule} implies a very subtle fine tuning between an infinite number
of couplings and masses of the hadronic spectrum and the anomaly. As we have seen, there
is no clear cut argument which would allow one to simply dismiss the presence of the free
quark $N_C/Q^2$ short distance contribution in $w_T(Q^2)$. However, from our experience
gained from the study of other QCD Green's functions, its persistence beyond the
perturbative regime leads to rather peculiar properties of $w_T(Q^2)$.

 We hope to have presented enough evidence that the combination
$w_L(Q^2) - 2\ w_T(Q^2)$, i.e. Eq. (\ref{eq:moment}), is a very interesting object for
the study of nonperturbative issues in QCD such as the spontaneous breakdown of chiral
symmetry, the Operator Product Expansion and the chiral anomaly. We hope that some of the
above observations will help attract further attention on the understanding of its
properties.

\vspace{1cm}

\noindent\textbf{Acknowledgements} \vspace{0.5cm}

\noindent We thank A. Vainshtein for discussions. We also thank A. Manohar, V. Miransky
and V. Zakharov for interesting correspondence. This work has been supported in part by
TMR, EC-Contract No. HPRN-CT-2002-00311 (EURIDICE). The work of S.P. has also been
partially supported by the research projects CICYT-FEDER-FPA2002-00748 and 2001-SGR00188.

 \vfill
\newpage

\noindent
{\Large\bf{Appendix A: perturbative proof of the
nonrenormalization theorems}}
\renewcommand{\theequation}{A.\arabic{equation}}
\setcounter{equation}{0}

\indent

\noindent
The perturbative expansion of ${\cal W}_{\mu\nu\rho}(q_1,q_2)$
is given by
\be
{\cal W}_{\mu\nu\rho}(q_1,q_2)
\vert_{\mbox{\scriptsize{pQCD}}}
 \,=\,
(-1)i^4\,\frac{N_C}{2} \sum_{n=0}^\infty
\left(\frac{\alpha_s}{\pi}\right)^n\,
{\cal W}_{\mu\nu\rho}^{[n]}(q_1,q_2)
\,.
\ee
The lowest order contribution arises from the free
quark triangle, and reads
\be
{\cal W}_{\mu\nu\rho}^{[0]}(q_1,q_2) \,=\,
\Gamma_{\mu\nu\rho}^{[0]}(q_1,q_2\vert a)\,+\,
\Gamma_{\nu\mu\rho}^{[0]}(q_2,q_1\vert b)
\,,
\ee
with
\be
\Gamma^{[0]}_{\mu\nu\rho}(q_1,q_2\vert a)\,=\,\int\,\frac{d^4P}{(2\pi)^4}\,
{\mbox{tr}}\bigg[\,\frac{1}{\not\!
 P-\not\! q_1+\not\! a-m_q}\,\gamma_\mu\,\frac{1}{\not\! P+\not\! a-m_q}\,\gamma_{\nu}\,
\frac{1}{\not\! P+\not\! q_2+\not\! a-m_q}\,\gamma_{\rho}\gamma_5\,\bigg]
\,.
\ee
In these expressions, the arbitrary four-vectors $a_\mu$
and $b_\mu$ parametrize the ambiguity in defining the
routing of the loop momentum $P_\mu$. This ambiguity
reads
\be
\Gamma^{[0]}_{\mu\nu\rho}(q_1,q_2\vert a) \,=\,
\Gamma^{[0]}_{\mu\nu\rho}(q_1,q_2)\,-\,
\frac{1}{8\pi^2}\,\epsilon_{\mu\nu\rho\sigma}a^{\sigma}
\,,
\ee
where $\Gamma^{[0]}_{\mu\nu\rho}(q_1,q_2)$ corresponds
to $\Gamma^{[0]}_{\mu\nu\rho}(q_1,q_2\vert a)$ with the
choice $a_\mu = (0,0,0,0)$. Therefore,
\be
{\cal W}_{\mu\nu\rho}^{[0]}(q_1,q_2) \,=\,
\Gamma_{\mu\nu\rho}^{[0]}(q_1,q_2)\,+\,
\Gamma_{\nu\mu\rho}^{[0]}(q_2,q_1)\,-\,
\frac{1}{8\pi^2}\,\epsilon_{\mu\nu\rho\sigma}(a-b)^{\sigma}
\,
\ee
The difference $(a-b)_\mu$ is fixed upon imposing the
conservation of the vector current. For $m_q=0$, one has
\bea
(q_1 + q_2)^\rho \Gamma_{\mu\nu\rho}^{[0]}(q_1,q_2)
&=& 0
\nonumber\\
q_1^\mu \left[
\Gamma_{\mu\nu\rho}^{[0]}(q_1,q_2)\,+\,
\Gamma_{\nu\mu\rho}^{[0]}(q_2,q_1)
\right] &=&
-\,\frac{1}{4\pi^2}\,\epsilon_{\mu\nu\rho\sigma}
q_1^\mu q_2^\sigma
\nonumber\\
q_2^\nu \left[
\Gamma_{\mu\nu\rho}^{[0]}(q_1,q_2)\,+\,
\Gamma_{\nu\mu\rho}^{[0]}(q_2,q_1)
\right] &=&
-\,\frac{1}{4\pi^2}\,\epsilon_{\mu\nu\rho\sigma}
q_1^\nu q_2^\sigma
\,.
\eea
Therefore,
$\{q_1^\mu ; q_2^\nu\}{\cal W}_{\mu\nu\rho}^{[0]}(q_1,q_2)
= \{0 ; 0\}$ provided $(a - b)_\mu = 2 (q_1 - q_2)_\mu$,
which, for massless quarks, then leads to
\be
(q_1 + q_2)^\rho {\cal W}_{\mu\nu\rho}^{[0]}(q_1,q_2)
\, = \, \frac{1}{2\pi^2}\,\epsilon_{\mu\nu\rho\sigma}
q_1^\rho q_2^\sigma
\,.
\ee

\indent

\noindent
Staying in the chiral limit, one has, upon
shifting the loop momentum
\bea
\Gamma_{\mu\nu\rho}^{[0]}(q_1,q_2) &=&
\Gamma_{\nu\rho\mu}^{[0]}(q_2,-q_1-q_2\vert q_2)
\nonumber\\
&=& \Gamma_{\nu\rho\mu}^{[0]}(q_2,-q_1-q_2) \, - \,
\frac{1}{8\pi^2}\,\epsilon_{\mu\nu\rho\sigma} q_2^\sigma \,. \eea Consequently,
\be\label{anine} {\cal W}_{\mu\nu\rho}^{[0]}(q_1,q_2) \,-\, {\cal
W}_{\rho\nu\mu}^{[0]}(-q_1-q_2,q_2) \,=\, \frac{1}{2\pi^2}\,\epsilon_{\mu\nu\rho\sigma}
q_2^\sigma \,. \ee

\indent

\noindent As far as the higher order contributions ${\cal
W}_{\mu\nu\rho}^{[n]}(q_1,q_2)$, $n \ge 1$, are concerned, they may be written as \be {\cal
W}_{\mu\nu\rho}^{[n]}(q_1,q_2) \,=\, \Gamma_{\mu\nu\rho}^{[n]}(q_1,q_2)\,+\,
\Gamma_{\nu\mu\rho}^{[n]}(q_2,q_1) \,. \ee The integration over the triangle loop is now
well defined, provided all the subgraphs arising from the QCD corrections (for instance
quark and gluon self energies) have been properly regularized, i.e. assuming one works in
the same conditions which allow to prove the Adler-Bardeen non renormalization theorem
for the anomalous part. Then all shifts in the momenta are allowed, and one has \be
\Gamma_{\mu\nu\rho}^{[n]}(q_1,q_2) \,=\, \Gamma_{\nu\rho\mu}^{[n]}(q_2,-q_1-q_2) \,. \ee
Therefore, \be\label{atwelve} {\cal W}_{\mu\nu\rho}^{[n]}(q_1,q_2) \,-\, {\cal
W}_{\rho\nu\mu}^{[n]}(-q_1-q_2,q_2) \,=\, 0 \,. \ee

\indent

\noindent Taking (\ref{anine},\ref{atwelve},\ref{omegaw}) and (\ref{shift}) leads to \be
{\widehat\Omega}_{\mu\nu\rho} \vert_{\mbox{\scriptsize{pQCD}}}
 \,=\, 0
\,. \ee This by itself does not prove, but is certainly compatible with the fact that
${\widehat\Omega}_{\mu\nu\rho}$ is an order parameter. However, it suffices to entail the
non renormalization theorems discussed before.

\vspace {1cm} \noindent
 {\Large\bf{Appendix B: The function $\Omega_{\mu\nu\rho}$ in the
one-family Standard Model.}}
\renewcommand{\theequation}{B.\arabic{equation}}
\setcounter{equation}{0} \vspace{0.5cm}

\noindent Let us consider the first family of quarks \emph{and} leptons in the Standard
Model. In this case, the currents appearing in the Green's function $\Omega_{\mu\nu\rho}$
in Eq. (\ref{eq:omegadef}) are the complete ones, i.e. including not only the $u,d$
quarks but also the electron and the neutrino. These currents are made up with the
generators of the $SU(2)_L\times SU(2)_R$ subgroup of $SU(3)_L\times SU(3)_R$ which gets
gauged when the electroweak interactions are turned on. Indeed, this is the Green's
function which contributes to physical observables such as the muon $g-2$
\footnote{Again, we are neglecting the flavor singlet component of the Z boson.}. In the
case where $q_1=k\pm q$, $q_2=-k$ and $k$ is small, keeping only terms linear in this
momentum $k$ and working to lowest order in the electroweak interactions, one finds that
\begin{eqnarray}\label{app1}
  \Omega_{\mu\nu\rho}(k\pm q,-k) &=& \frac{1}{32 \pi^2}\ \biggl\{w^{TOTAL}_L(Q^2)
  \Bigl(-q_{\rho}\ \epsilon_{\mu\nu\alpha\sigma}\ q^{\alpha} k^{\sigma}- q_{\mu}\
  \epsilon_{\nu\rho\alpha\sigma}\ q^{\alpha} k^{\sigma}\Bigr) + \nonumber \\
   &+& 2\ w_T^{TOTAL}(Q^2) \left(-q^2\ \epsilon_{\mu\nu\rho\sigma}\ k^{\sigma}+
   q_{\mu}\ \epsilon_{\nu\rho\alpha\sigma}\ q^{\alpha} k^{\sigma}+ q_{\rho}\
  \epsilon_{\mu\nu\alpha\sigma}\ q^{\alpha} k^{\sigma}\right)\biggr\}\ .
\end{eqnarray}
However, in this case the functions $w^{TOTAL}_{L,T}(Q^2)$ contain quarks as well as
leptons. Therefore,
\begin{equation}\label{app0}
w^{TOTAL}_L(Q^2)= w^{quarks}_L(Q^2)+ w^{leptons}_L(Q^2)=0\ ,
\end{equation}
as a consequence of the anomaly cancellation in the Standard Model. For the function
$w^{TOTAL}_T$ one obtains
\begin{equation}\label{app2}
    2\ w^{TOTAL}_T(Q^2)=  2\ w^{quarks}_T(Q^2) + 2\ w^{leptons}_T(Q^2) \ ,
\end{equation}
but, obviously,  $w_L^{leptons}(Q^2)=2\ w_T^{leptons}(Q^2)$ to all orders in $\alpha_s$
because leptons do not experience strong interactions. Consequently,
\begin{equation}\label{app3}
    2\ w^{TOTAL}_T(Q^2)= 2\ w^{quarks}_T(Q^2) +  w^{leptons}_L(Q^2)\ ,
\end{equation}
and using Eq. (\ref{app0}), one finally obtains
\begin{equation}\label{app4}
    2\ w^{TOTAL}_T(Q^2) = 2\ w^{quarks}_T(Q^2) - w^{quarks}_L(Q^2)\ ,
\end{equation}
which, together with Eq. (\ref{app1}), is the result in Eq. (\ref{omegahat}) in the text.

\end{document}